\documentclass[prd,superscriptaddress,showpacs,floats,nofootinbib,preprint,floatfix]{revtex4}

\usepackage{graphicx}
\usepackage{epsfig}
\usepackage{relsize}  
\usepackage{bm}       
\usepackage{amsfonts} 
\usepackage{amsmath}  
\usepackage{amssymb}  
\usepackage{upgreek}  
\usepackage{dsfont}   
\usepackage{xspace}   
\usepackage{array}
\usepackage[font=footnotesize,bf,labelsep=space]{caption}
\usepackage{slashed}  

\newcommand{\be}{\begin{eqnarray}}
\newcommand{\ee}{\end{eqnarray}}
\newcommand{\ben}{\begin{eqnarray*}}
\newcommand{\een}{\end{eqnarray*}}
\newcommand{\dd}{{\mathrm{d}}}
\newcommand{\tr}{{\mathrm{Tr }}}

\newcommand{\esp}{{\mathrm e}}
\newcommand{\mref}[1]{(\ref{#1})}

\begin{document}

\author{W.M. Alberico} 
\author{S. Chiacchiera} 
\author{H. Hansen} 
\author{A. Molinari}
\author{M. Nardi}

\affiliation{Istituto Nazionale di Fisica Nucleare, Sezione di Torino\\ and
  Dipartimento di Fisica Teorica, via Giuria 1, I-10125 Torino, Italy}

\title{Shear Viscosity of Quark Matter}

\date{\today}

\begin{abstract}

We consider the shear viscosity of a system of quarks  and its ratio to the entropy 
density above the critical temperature for deconfinement. Both quantities are derived 
and computed for different modeling of the quark self-energy, also allowing for a 
temperature dependence of the effective mass and width. 
The behaviour of the viscosity and the entropy density is argued in terms of the 
strength of the coupling and of the main characteristics of the quark self-energy. 
A comparison with existing results is also discussed.

\end{abstract}

\maketitle
\section{Introduction \label{introduction}}

We address the problem of the viscosity of a Fermi system, in particular of quark matter. Much interest on this physics has been stimulated by the experiments carried out at the Relativistic Heavy Ion Collider (RHIC): these seem to indicate that above the critical temperature $T_C$ the quark-gluon plasma (QGP) behaves as a fluid with a quite small viscosity (i.e. as an almost perfect fluid). 
This conclusion~\cite{Huovinen,Drescher07,Csernai}, mostly based on the findings on 
the $v_2$ coefficient measured~\cite{Phenix1,Star,Phobos,Phenix2} in the multipole 
analysis of the angular distribution of the hadrons produced in an ultrarelativistic 
ion-ion collision, contrasts with a description of the QGP as a fluid of almost 
independent Landau quasi-particles which emerges from lattice 
calculations~\cite{Karsch} though more recent findings~\cite{Meyer} substantially 
lowered also the lattice predictions.
Also the substantial collective flow (the ``elliptic'' flow) observed in these collisions appears to imply that the viscosity can not be that 
large\cite{Teaney-shuryak,Teaney-03, Teaney-04,Peshier,KapustaQM}.
In this work we shall examine this challenging question, confining ourselves, at present, to the quark degrees of freedom only~: hence our results will not be directly comparable with the indications of the RHIC data, but should be viewed as first a step in that direction, in the absence of the gluonic degrees of freedom.

We will consider, beyond the shear viscosity $\eta$, also the ratio $\eta/s$, $s$ being the entropy density of the system. Indeed  $\eta/s$, which for vanishing chemical potential 
basically governs the damping rate of sound waves propagating into the 
system~\cite{Teaney-03}, has much physical significance: its temperature behaviour should 
determine the value $T_C$ at which a phase transition occurs in the 
system~\cite{Csernai,Lacey}.

It should also be added that certain special supersymmetric $N=4$ field theories (dual to black branes in higher space-time dimensions) predict a lower limit for the viscosity/entropy density ratio, namely $\eta/s\geqslant 1/4\pi$ (in units $\hbar=k_B=c=1$)~\cite{Kovtun}.

After introducing, in Section~\ref{formalism}, the basic formalism to derive the shear-viscosity and the entropy density starting from a dressed quark propagator (and the 
corresponding spectral function) we present in Section~\ref{results} our results and 
concluding remarks.


\section{Formalism \label{formalism}}
We base our investigation, along with many 
authors~\cite{Kovtun,Iwasaki,Fernandez-Fraile,Jeon,ValleBasagoiti}, on the linear response theory which, as far as 
the viscosity is concerned, leads to the Kubo formula~:
\be
\eta(\omega)=\frac{i}{\omega}\left[\Pi^R(\omega)-\Pi^R(0)\right]
\label{eq:Kubo}
\ee
where the retarded Green's function at zero momentum is defined as
\be
\Pi^R(\omega)=-i\int_0^\infty  \dd t\, \esp^{i\omega t} \int d^3r
<[T_{xy}({\bf r},t),T_{xy}(0,0)]>
\label{eq:retard}
\ee
and the integrand is  the correlator of the off-diagonal $(x,y)$ element of the energy-momentum tensor $T_{\mu\nu}$ at different space-time points.
In Eq.~(\ref{eq:retard}) the brackets imply the thermal average of the commutator. 

From \mref{eq:Kubo} and \mref{eq:retard} the static $(\omega=0)$ viscosity follows:
\be
\eta\equiv \eta(\omega=0)= - \left. \frac{\dd}{\dd \omega}
{\mathrm{Im}}\, \Pi^R(\omega) \right|_{\omega=0^+}~.
\ee

We conveniently get $\Pi^R$ from the Matsubara formalism by analytic continuation,
 according to the prescription ($\delta\to 0$)
\ben
\Pi^R(\omega)=\left. \Pi(i\omega_n) \right|_{i\omega_n=\omega+i\delta}
\een
with
\be
\Pi(i\omega_n)= -\int_0^\beta \dd\tau\, \esp^{-i\omega_n\tau}
\int  d^3r <T_\tau\Big(T_{xy}({\bf r},\tau) T_{xy}(0,0)\Big)>\, ;
\label{eq:Pi}
\ee
in the above $\omega_n=2\pi n/\beta$, $\beta=1/T$ and $T_\tau$ is the $\tau$-ordered product.

Concerning the structure of the canonical energy-momentum tensor we observe that for all the Lagrangians not displaying a derivative coupling among the fields (a typical example being the Nambu--Jona-Lasinio one) it turns out that
\be
T_{xy}=\frac{i}{2}\left(\bar{\psi}\gamma_2\partial_1 \psi -
\partial_1\bar{\psi}\gamma_2 \psi\right)~, \label{eq:Jxy}
\ee
$\psi$ being the quark field.

Following Ref.~\cite{Iwasaki}, in the evaluation of $\Pi(i\omega_n)$ one can
stick to the first order of a ring diagram expansion, since higher orders (e.g. for a scalar or pseudoscalar interaction coupling) vanish, in the chiral limit $m_q=0$, 
due to the trace of odd numbers of 
$\gamma$ matrices\footnote{A different situation would be encountered by considering 
series of ladder diagrams~\cite{Fernandez-Fraile}; however one should keep in mind that 
exchange-like diagrams (interaction inside a fermionic loop) are reduced with respect to the direct ones by the degeneracy of states, including the number of colors $N_c$.}. Hence, by inserting \mref{eq:Jxy} into \mref{eq:Pi}, one easily gets:
\be
\Pi(i\omega_n)=\frac{1}{\beta} \sum_l
\int \frac{d^3 p}{(2\pi)^3}\, p^2_x \,\tr \left[
\gamma_2 S(i\omega_l+i\omega_n,{\bf p})
\gamma_2 S(i\omega_l,{\bf p}) \right]
\ee
where the trace is taken over spin, flavor and colour.

Next, by introducing  the spectral representation of the quark propagator, 
\be
S(i\omega_l,{\bf p})=\int_{-\infty}^\infty
\frac{\dd \epsilon}{2\pi} \, 
\frac{\rho(\epsilon,{\bf p})}
{i\omega_l-\epsilon}\, ,
\ee
with $\omega_l=(2l+1)\pi/\beta$, it is possible~\cite{Iwasaki} to carry out the summation over the Matsubara frequencies via standard contour integral technique. 
One ends up with the following expression for the shear viscosity:
\be
\eta=-\frac{1}{2} \int_{-\infty}^\infty
\frac{\dd \epsilon}{2\pi} \, 
\int \frac{d^3 p}{(2\pi)^3}\, p^2_x \,
\frac{\partial f}{\partial \epsilon} \,
\tr\left[ \rho(\epsilon,{\bf p})\gamma_2
\rho(\epsilon,{\bf p})\gamma_2 \right]\label{eq:eta}
\ee
where $f(\epsilon)=1/(\esp^{\beta(\epsilon-\mu)}+1)$ is the thermal distribution
for fermions; we notice that the viscosity $\eta$ gets its major contribution from the surface of the Fermi distribution.

In equation \mref{eq:eta} the quark propagators are meant to be ``fully'' 
dressed within the appropriate mean-field approach, depending upon the
model Lagrangian one is referring to. Here we shall adopt a merely 
phenomenological approach, with a suitable Ansatz for the quark self-energy. 

\subsection{The scalar case}
We consider first the case of a Lorentz scalar self-energy of the form
\be
\Sigma(p)=\left[M(p)-i\Gamma(p)\right] {\mathds 1}
\label{eq:sigmasca}
\ee
to be inserted into the (massless) fermion propagator 

\be
S=\frac{1}{\slashed{p}-\Sigma(p)}~. \label{eq:propS}
\ee
In the above, $M$ and $\Gamma$ should be viewed as phenomenological 
functions of the four-momentum $p$, which will be specified later on.

The spectral function associated with the propagator \mref{eq:propS}
is then easily found to be

\be
\rho(\epsilon,{\bf p})= -i\,
\left[ S_A(\epsilon,{\bf p})-S_R(\epsilon,{\bf p})\right]
= 2 \frac{\Gamma\,  \mbox{sign}(\epsilon)\left((\slashed{p}+M)^2+\Gamma^2\right)}
{\left({p}^2-M^2+\Gamma^2\right)^2+4M^2\Gamma^2}
\label{eq:spectrsca}
\ee
where $S_A$ and $S_R$ are the usual advanced and retarded propagators.
\footnote{Notice that (\ref{eq:spectrsca}) differs from the Lorentzian shape one would obtain in the non-relativistic limit.}

After inserting \mref{eq:spectrsca} into \mref{eq:eta} and performing the relevant 
traces one gets the following expression for the viscosity:
\label{eq:etaIeII}
\be
\eta = \eta_I+\eta_{II}
\ee
with
\begin{subequations}
\be\label{eq:etaI}
\eta_I=\frac{64 N_c N_f}{T} \int \frac{d^3 p}{(2\pi)^3}
\int_{-\infty}^\infty \frac{\dd \epsilon}{2\pi}
\left(1-f(\epsilon)\right) \, f(\epsilon)\, M^2\Gamma^2
\frac{p_x^2p_y^2}{\left[(p^2-M^2+\Gamma^2)^2+4M^2\Gamma^2\right]^2}
\ee
and
\be\label{eq:etaII}
\eta_{II}= -\,\frac{8N_cN_f}{T}  \int \frac{d^3 p}{(2\pi)^3}
\int_{-\infty}^\infty \frac{\dd \epsilon}{2\pi}
\left(1-f(\epsilon)\right) \, f(\epsilon)\, \Gamma^2
\frac{p_x^2}{(p^2-M^2+\Gamma^2)^2+4M^2\Gamma^2}~.
\ee
\end{subequations}
In the above $p^2=\epsilon^2-{\bf p}^2$, $N_c$ and $N_f$ are the
colour and flavour numbers. The temperature (and chemical potential)  
dependence is embedded into  the Fermi distribution $f(\epsilon)$.
In the similar approach by Iwasaki et al.~\cite{Iwasaki} $M$ and $\Gamma$ 
were kept as positive, constant parameters; however with this choice $\eta_{II}$,
although generally smaller than $\eta_{I}$ if integrated up to a cutoff momentum,
 is divergent. 

In order to ensure the convergence of $\eta_{II}$ and guided by simplicity 
arguments,  we shall consider a constant parameter $M$, while for the 
width of the quasi-particle we use:
\be\label{eq:lambda}
\Gamma(p) = \frac{\lambda^2}{\sqrt{|{\bf p}|^2+M^2}}\,,
\ee
$\lambda$ being a constant parameter as well. One should notice that the choice of 
the two parameters entering into the quasi-particle self-energy is not completely arbitrary, since on the basis of general arguments the spectral function must obey the following "sum rule"~\cite{Weldon00}:
\be\label{eq:sumrule}
\frac{1}{4}\tr_{\mathrm{Spin}}\int_{-\infty}^{\infty}\frac{d\epsilon}{2\pi} 
\left[ \rho(\epsilon,p)\gamma_0\right] = 1\, ,
\ee
which is satisfied when the fermion propagator obeys a dispersion relation.

Let us now consider the entropy density within the same model (we stick in this paper 
to the $\mu=0$ case); according to the customary field theory 
formulation~\cite{Kapusta-book} it reads:
\be
s&&=\frac{1}{V}\frac{\partial}{\partial T}\left(T\ln Z\right)
\label{scalarentropyI}\\
&&= 2 N_c N_f \int\frac{d^3 p}{(2\pi)^3}\left\{\ln\left(1+e^{-\beta\omega_+}\right)
+ \ln\left(1+e^{-\beta\omega_-}\right)+\frac{\beta\omega_+}{1+e^{\beta\omega_+}}
+ \frac{\beta\omega_-}{1+e^{\beta\omega_-}} \right\}\nonumber
\ee
where $V$ is the normalization volume and 
\be
\omega_{\pm}^2(p) = {\bf p}^2+M^2-\Gamma^2\pm 2i\Gamma M\,.
\ee
By separating real and imaginary parts of $\omega_{\pm}(p)$ it can be explicitly 
shown that $s$ is real; one gets (in the hypothesis $\Gamma < M$):
\be
s= 2 N_c N_f \!\!\int\!\!\frac{d^3 p}{(2\pi)^3}\left\{\!\ln\left[e^{-2{\widetilde E}_p} +
2e^{-{\widetilde E}_p }\cos{{\widetilde\alpha}_p}+1\right]\! + \!
2e^{-{\widetilde E}_p }\frac{{\widetilde E}_p\left(e^{-{\widetilde E}_p }+
\cos{{\widetilde\alpha}_p}\right) + {\widetilde\alpha}_p\sin{{\widetilde\alpha}_p}}
{e^{-2{\widetilde E}_p} + 2e^{-{\widetilde E}_p }\cos{{\widetilde\alpha}_p}+1}\!
\right\}
\label{scalarentropyII}
\ee
with ${\widetilde E}_p=\beta\,{\mathrm {Re}}~\omega_+(p)$, 
${\widetilde\alpha}_p=\beta\,{\mathrm {Im}}~\omega_+(p)$.

\subsection{General self-energy}
Let us now turn to a more general structure for the quark self-energy:
\be
\Sigma(p)=a_0(p)\gamma^0+a_1(p)\,{\bm \upgamma}\cdot{\bf p} +
a_2(p){\mathds 1} \label{eq:SigmaLo}
\ee
$p$ being the quark four-momentum. The corresponding propagator and spectral function 
can be more conveniently expressed making use of the 
customary projector operators~\cite{Blaz-olli,BIR3}: 
\be
\Lambda_{\pm}({\bf p})= \frac{1}{2}\left({\mathds 1}\pm 
\gamma_0\frac{{\bm \upgamma}\cdot{\bf p}+m}{E_p}\right)
\label{eq:project}
\ee
where $E_p=\sqrt{{\bf p}^2+m^2}$ and $m$ is the bare fermion mass.

It is then easily shown that:
\be
\gamma_0 S^{-1}(p)\equiv \gamma_0 \left({\slashed{p}-m-\Sigma(p)}\right)=
\Delta^{-1}_{+}(p)\,\Lambda_{+}({\bf p})+ \Delta^{-1}_{-}(p)\,\Lambda_{-}({\bf p})\, ,
\label{eq:propSLo}
\ee
with
\be
\Delta^{-1}_{\pm}(p_0,{\bf p})=p_0\mp [E_p+\Sigma_{\pm}(p)]\,.
\label{eq:deltapm}
\ee
In the above the self-energy (\ref{eq:SigmaLo}) has been expressed as well in terms of the 
projection operators (\ref{eq:project}), with the condition $a_2(p)=ma_1(p)$:
\be
\gamma_0 \Sigma(p)= \Sigma_{+}(p)\,\Lambda_{+}({\bf p})- 
\Sigma_{-}(p)\,\Lambda_{-}({\bf p})\, ,
\label{eq:SigmaLoproj}
\ee
where
\be
 \Sigma_{\pm}(p)= a_1(p) E_p \pm a_0(p)\, .
\label{eq:Sigmapm}
\ee
The functions $a_i(p)\, (i=0,1)$ should be obtained on the basis of some microscopic 
calculation, thus entailing the modeling of the quark propagator in a suitable 
description.

Attention should be payed to the definition of the retarded and advanced propagator, 
for which analogous definitions hold, e.g. with
\be
\Delta^{R(A)}_{\pm}(p_0,{\bf p})=\frac{1}{p_0\mp [E_p+\Sigma^{R(A)}_{\pm}(p)]}\,.
\label{eq:deltapmR}
\ee
By definition the ``retarded'' self-energy must satisfy the condition:
\be
{\mathrm {Im}}\Sigma^{R}_{+}< 0 ~~~~~{\mathrm {and}}~~~~~ {\mathrm {Im}}\Sigma^{R}_{-}> 0
\ee
and
\be
\Sigma^{A}_{\pm}(p)= \left[\Sigma^{R}_{\pm}(p)\right]^*\,.
\ee

The spectral function can now be expressed as
\be
\rho(p_0,{\bf p})\gamma_0 = \rho_{+}(p_0,{\bf p})\,\Lambda_{+}({\bf p}) + 
 \rho_{-}(p_0,{\bf p})\,\Lambda_{-}({\bf p})
\label{rhoLoproj}
\ee
where
\be
 \rho_{\pm}(p_0,{\bf p}) = \mp\frac{2{\mathrm {Im}}\Sigma^{R}_{\pm}(p)}
{\left[p_0\mp\left(E_p+{\mathrm {Re}}\Sigma^{R}_{\pm}(p)\right)\right]^2 + 
\left[ {\mathrm {Im}}\Sigma^{R}_{\pm}(p)\right]^2}\,.
\label{rhopmLoproj}
\ee

With these ingredients we can now turn to the evaluation of the shear viscosity and of 
the entropy density; the former is obtained from formula (\ref{eq:eta}) and again can 
be split into two terms, $\eta_I$ and $\eta_{II}$, which read:

\begin{subequations}
\be\label{eq:etaILo}
\eta_I=\frac{ N_c N_f}{T} \int \frac{d^3 p}{(2\pi)^3}
\int_{-\infty}^\infty \frac{\dd \epsilon}{2\pi}
\left(1-f(\epsilon)\right) \, f(\epsilon)\, \frac{p_x^2p_y^2}{E_p^2}\left[
\rho_{+}(\epsilon, {\bf p}) - \rho_{-}(\epsilon, {\bf p})\right]^2
\ee
and
\be\label{eq:etaIILo}
\eta_{II}= \frac{N_cN_f}{T}  \int \frac{d^3 p}{(2\pi)^3}
\int_{-\infty}^\infty \frac{\dd \epsilon}{2\pi}
\left(1-f(\epsilon)\right) \, f(\epsilon)\,
{2p_x^2}\rho_{+}(\epsilon, {\bf p})\rho_{-}(\epsilon, {\bf p})~.
\ee
\end{subequations}

For what concerns the entropy density, we start from the general formula~\cite{BIR3}
\be
s=-2N_cN_f \int \frac{d^3 p}{(2\pi)^3}\,\int_{-\infty}^{\infty} \frac{d \epsilon}{(2\pi)} 
\frac{\partial  f(\epsilon)}{\partial T}\, \tr_{\mathrm{Spin}} \left\{{\mathrm {Im}} \left[ 
\ln \left(-\gamma_0S^{-1}_R\right)\right] + {\mathrm {Im}}\left(\gamma_0\Sigma^R\right)
{\mathrm {Re}}\left(S_R\gamma_0\right)\right\}\,,
\label{eq:entropyLo}
\ee
the trace being taken over the spin degrees of freedom. Having expressed all relevant 
quantities in terms of the projector operators, the trace of the logarithm can now be 
 carried out, yielding
\be
s&&=-4N_cN_f \int \frac{d^3 p}{(2\pi)^3}\,\int_{-\infty}^{\infty} \frac{d \epsilon}{(2\pi)} 
\frac{\partial  f(\epsilon)}{\partial T}\,\left\{\mbox{arctg}\left(\frac{-{\mathrm {Im}}\Sigma^R_+}
{\epsilon-(E_p+{\mathrm {Re}}\Sigma^R_+)}\right)+\right.
\nonumber\\
&&+\mbox{arctg}\left(\frac{{\mathrm {Im}}\Sigma^R_-}{\epsilon+E_p+{\mathrm {Re}}\Sigma^R_-}\right)
-\pi\theta[\epsilon-(E_p+{\mathrm {Re}}\Sigma^R_+)]
-\pi\theta[\epsilon+E_p+{\mathrm {Re}}\Sigma^R_-]+
\nonumber\\
&&\left. -\frac{1}{2}\rho_+(\epsilon,p)[\epsilon-(E_p+{\mathrm {Re}}\Sigma^R_+)]-
\frac{1}{2}\rho_-(\epsilon,p)[\epsilon+E_p+{\mathrm {Re}}\Sigma^R_-]\right\}\,.
\label{eq:entropyLo1}
\ee

We now apply the above formalism to a schematic model for the quark self-energy. 
We stick to the chiral limit ($m=0$) and start
from the HTL self-energy~\cite{BIR2,BIR3} ($p\equiv|{\bf p}|$):
\be
\Sigma^{HTL}_{\pm}(\epsilon,p)=\frac{M^2}{p}\left(1-\frac{\epsilon\mp p}{2p}
\ln\frac{\epsilon+p}{\epsilon-p}\right)\,;
\label{selfHTL}
\ee
then, by taking into account that at $\mu=0$ the integrand in the shear viscosity is 
peaked at zero frequency, we consider the $\epsilon\to 0^+$ limit (namely 
$\epsilon\to i\delta$, $\delta$ being a positive infinitesimal) of the expression
(\ref{selfHTL}):
\be
{\hat{\Sigma}_\pm}^R(p)\equiv \Sigma^{HTL}_{\pm}(0^+,p) =
\frac{M^2}{p}\mp i\pi\frac{M^2}{2p}\,.
\label{selfour}
\ee
In the above formulas the temperature dependent mass parameter is $M=\lambda T$, where
(in the original HTL formula) $\lambda^2=C_f g^2/8=g^2/6$ for $N_c=3$ colors, $C_f$ being the Casimir invariant of $SU(N_c)$ and $g$ the strong coupling constant.

The shear viscosity is then obtained by inserting into eqs.~(\ref{eq:etaILo}) and 
(\ref{eq:etaIILo}) the following spectral densities:
\be
\hat{\rho}_\pm (\epsilon,p)=\frac{\pi M^2}{p}\frac{1}
{\left[\epsilon\mp \left(p+\frac{M^2}{p}\right)\right]^2+\left(\frac{\pi M^2}{2p}\right)^2}\,.
\label{rhoour}
\ee

In order to evaluate the entropy density (\ref{eq:entropyLo1}) one can conveniently 
employ the identity
\be
\frac{\partial f(\epsilon)}{\partial T}=-\frac{\partial\sigma_f(\epsilon)}
{\partial \epsilon}\,,
\ee
with
\be
\sigma_f(\epsilon)=-\left[f(\epsilon)\log f(\epsilon)
+(1-f(\epsilon))\log(1-f(\epsilon))\right]\,;
\ee
it allows to integrate by parts over the energy in (\ref{eq:entropyLo1}), yielding 
the simple expression\footnote{The finite term vanishes since $\lim_{\epsilon\rightarrow\pm\infty}\sigma_f(\epsilon)=0$.} 
\be
s=\pi N_c N_f \int \frac{d^3 p}{(2\pi)^3}\,\int_{-\infty}^{\infty} \frac{d \epsilon}{(2\pi)} \,\sigma_f(\epsilon)\,\frac{M^2}{p}
\left\{\hat{\rho}_+^2 (\epsilon,p)+\hat{\rho}_-^2 (\epsilon,p)\right\}\,.
\label{eq:entropyour2}
\ee

The function $\pi\frac{M^2}{p}\left\{\hat{\rho}_+^2 (\epsilon,p)+
\hat{\rho}_-^2 (\epsilon,p)\right\}$ can be interpreted as the spectral 
density~\cite{Baym} for the
entropy. Notice that in getting (\ref{eq:entropyour2})  the energy independence of the 
self-energy (\ref{selfour}) was crucial; moreover we removed the contributions stemming from the derivative of the theta-functions in Eq.~(\ref{eq:entropyLo1}), since they 
would have produced extra delta-peaks in the spectral density, leading to a double 
counting of the degrees of freedom (and altering the normalization of the spectral 
function itself).

\section{Results and conclusions \label{results}}

Using the formalism of the previous Section, we have evaluated the shear viscosity, 
the entropy density and their ratio with three different choices for the quark self-energy.
For the simplest scalar self-energy [Eq.~(\ref{eq:sigmasca})] we first consider a constant $M$ and $\Gamma$ given by Eq.~(\ref{eq:lambda}) with $\lambda=0.5 M$; the choice for the latter was constrained by the sum rule for the spectral density, which, for  $\lambda\le 0.5 M$ is only mildly violated (e.g. with the above value it differs from 1 by less than about $10\%$ in the interval  $0\le p \le 150$~MeV).

\begin{figure}[h]
  \begin{center}
    \epsfig{file=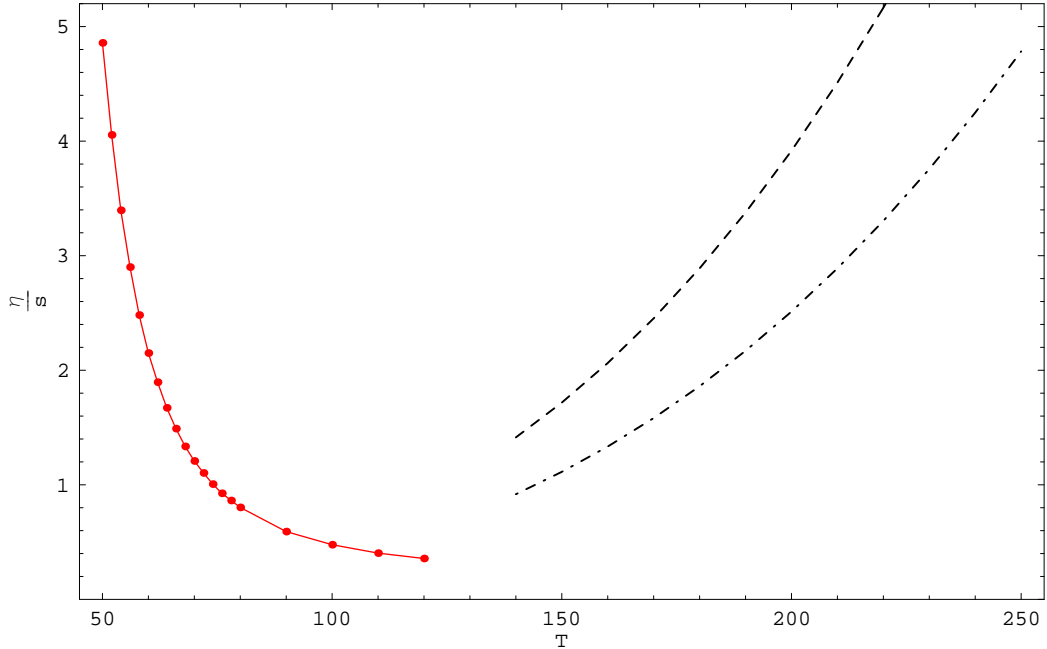,width=0.85\textwidth}
    \caption{The ratio $\eta/s$ in the scalar model with $M=300$~MeV (dashed line) 
and $M=350$~MeV (dot-dashed line) as a function of temperature for $T>140$~MeV; the 
points at low temperatures refer to the pion gas (Courtesy of D. Davesne~\cite{Dany}). }
    \label{fig:F01}
  \end{center}
\end{figure}

In Fig.~\ref{fig:F01} we display the result obtained for $\eta/s$ with $M=300$~MeV and 
$350$~MeV, as a function of temperature for $T\ge 140$~MeV: the ratio grows like $aT^3$, the coefficient $a$ becoming smaller as $M$ increases. We expect this model to be meaningful only in the regime of high temperatures, where it represents a gas of 
quasi-particles. It might be interesting to compare these results with the corresponding ratio obtained in a pion gas~\cite{Dany}, also displayed in the figure.
In this approach the shear viscosity  is derived from a Boltzmann-Uehling-Uhlenbeck transport equation via the Chapman-Enskog method~\cite{Davesne} to first order, while the entropy density is the local equilibrium bosonic entropy density.
Since the differential cross section used in the numerical computation is the 
experimental one, these values of $\eta/s$ at zero chemical potential are essentially 
model-independent. The two regimes (composite hadrons at low temperature, gas of 
quasi-particle quark states in the high $T$ bath) qualitatively reproduce the analogous situation illustrated  in Ref.~\cite{Csernai}, where the quark-gluon phase is described within perturbative QCD.

\begin{figure}[h]
  \begin{center}
    \epsfig{file=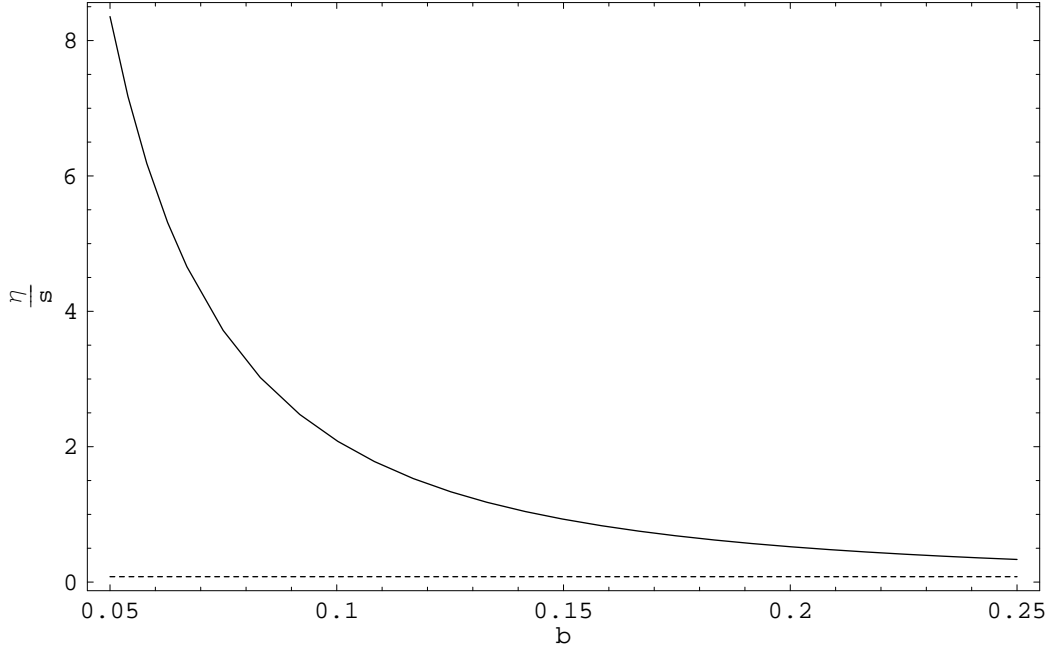,width=0.85\textwidth}
    \caption{Ratio $\eta/s$ in the scalar model with temperature dependent mass and 
width as a function of $b\equiv \lambda/M$. The AdS/CFT limit is reported (dashed line).}
    \label{fig:F02}
  \end{center}
\end{figure}

Within the same model, but taking into account a temperature dependence of the quasi-particle mass and width, 
the results for the ratio $\eta/s$ drastically change: indeed 
by heuristically assuming the conventional thermal free mass, $M=2\pi T$ and, in 
eq.(\ref{eq:lambda}), $\lambda= b M$, we obtained a  $T$-independent $\eta/s$ ratio, 
which displays a $1/b^2$ behaviour, as it is illustrated in Fig.~\ref{fig:F02}. 
We notice that for the largest value of $b$ reported in the figure, our $\eta/s$ ratio 
is close to the AdS/CFT limit; moreover for $b>0.25$ the shear viscosity becomes 
negative, an outcome which reflects the violation of the unitarity condition imposed 
by the sum rule (\ref{eq:sumrule}); the latter indeed "protects" the positivity of 
$\eta$ and the balance between Re$\Sigma$ and Im$\Sigma$: both grow linearly with $T$, 
but their ratio must be kept within well defined limits (with $\lambda=0.25 M$ the 
 violation of the sum rule is smaller that $5\%$).

\begin{figure}[h]
  \begin{center}
    \epsfig{file=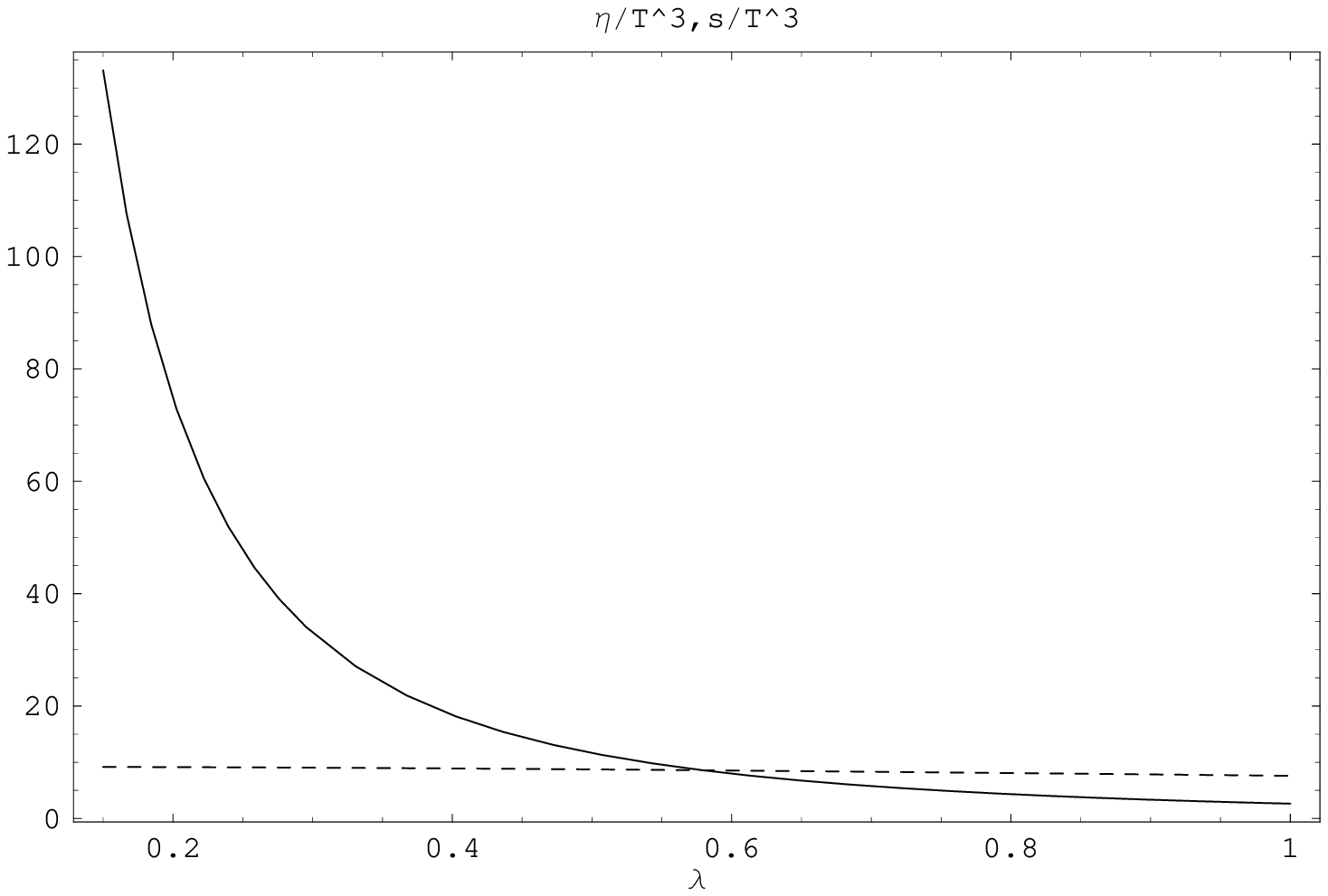,width=0.85\textwidth}
    \caption{The behaviour of  $\eta/T^3$ (continuous line) and $s/T^3$ (dashed line) 
versus $\lambda$ obtained with the self-energy (\ref{selfour}).}
    \label{fig:F03}
  \end{center}
\end{figure}

\begin{figure}[h]
  \begin{center}
    \epsfig{file=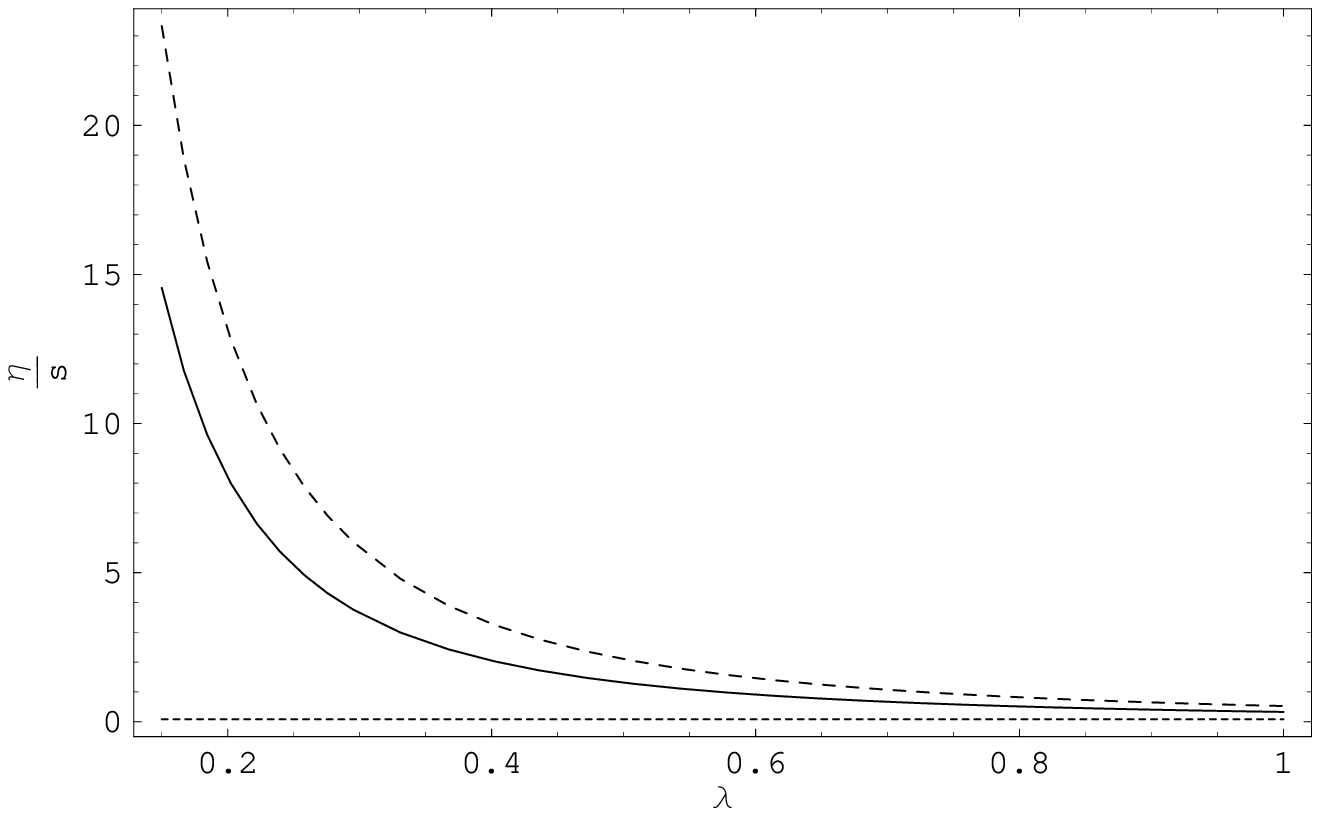,width=0.85\textwidth}
    \caption{The ratio $\eta/s$ versus $\lambda$ as obtained with the self-energy 
(\ref{selfour}) (continuous line) and with the scalar, T-dependent model (dashed line); 
 The AdS/CFT limit is also reported.}
    \label{fig:F04}
  \end{center}
\end{figure}

Finally we considered the self-energy of Eq.~(\ref{selfour}), which exactly satisfies 
the sum rule (\ref{eq:sumrule}) for any choice of $M$. As in the previous case, we 
found that both the shear viscosity and the entropy density raise proportionally to 
$T^3$, so that their ratio is temperature independent. This can be seen by
dimensional analysis since, sticking to $\mu=0$ and with a fixed strong
coupling constant (namely a T-independent $g$), the energy scale is set
by the temperature itself.

In Fig.~\ref{fig:F03} we report the quantity $\eta/T^3$ from 
Eqs.~(\ref{eq:etaILo}), (\ref{eq:etaIILo}) and 
$s/T^3$ from Eq.~(\ref{eq:entropyour2}) as a function of $\lambda$, while
Fig.~\ref{fig:F04} shows the corresponding ratio $\eta/s$: the latter is compared 
with the same quantity obtained in the scalar 
case, with a temperature dependent mass, the $b$ and $\lambda$ parameters being linked 
(at $T=0$) by the relation $b=\lambda/\sqrt{8\pi}$. 
We find that in both cases $\eta/s$ decreases as $1/\lambda^2$ and (since the
entropy density is slightly varying with $\lambda$) the dominant trend is due
to $\eta$. This $1/\Gamma$ behaviour ($\Gamma$ being the particle width) is a
general feature of the microscopic calculation of transport coefficients, which stems 
from the analytic properties of the fermion spectral function~\cite{Fernandez-Fraile}; likewise in kinetic theory the same behaviour is expected since $1/\Gamma$ is related 
to the mean free path of particles. 

Also quantitatively the results we find are compatible with previous ones.
For example at $\lambda=0.65$  ($g\simeq 1.6$) we find $\eta/T^3=6.7$ to
be compared with $\eta/T^3\simeq 7$ found by \cite{Zhuang} for a quark
plasma near the critical temperature. For the same $\lambda$ we get
$\eta/s\simeq 0.8$ which is four times larger than the one obtained in 
Ref.~\cite{Peshier} for a purely gluonic system in a similar approach. 
Even by pushing our model to $\lambda=1$ we would find the somewhat larger value 
$\eta/s\simeq 0.4$: it seems that in the present approximation the quark contribution
to $\eta/s$ is  larger than the gluonic one (at least if we restrict ourselves to
values of $\lambda$ for which the "HTL" approximation can be reasonably applied).

In spite of the different Lorentz structure of the self-energy the two ratios 
displayed in Fig.~\ref{fig:F04} are quite similar.
We can thus infer that the relevant feature which governs the 
ratio $\eta/s$ in the fermionic system is the temperature dependence of the effective 
mass and width of the quark, while the detailed (Lorentz) structure of the self-energy 
does not appear to be of much relevance.

In summary, we have considered in this paper the $\eta/s$ ratio for a system of quarks 
with the purpose of investigating the impact on this quantity of the effective quark 
self-energy, which, although modeled on the basis of simplicity, still grasped important 
aspects of a realistic description. 
All calculations were carried out at zero chemical potential. Only 
for the case of a constant mass parameter we find that $\eta/s$ grows with the 
temperature, as expected for a limiting case of a dilute gas system~\cite{Csernai,Lacey}.

 On the contrary, 
by assuming mass and width parameters proportional to the temperature, no matter 
upon the details of the self-energy, the ratio $\eta/s$ becomes temperature independent, 
since both $\eta$ and $s$ acquire a $T^3$ behaviour. 

Concerning the specific values obtained for this ratio, in our models it obviously 
depends on the coupling parameter ($b$ or $\lambda$) adopted in the definition of the 
quark self-energy; we found a general $1/\lambda^2$ behaviour of $\eta/s$, in agreement 
with previous findings. Our results are compatible with very small values of the ratio, 
but the fundamental requirement related to the sum rule obeyed by the quark spectral 
function sets some intrinsic limits on the value of the parameter itself and the 
AdS/CFT boundary is not reached. 

In the present paper we have shown that, within relatively simple models, one can 
obtain a realistic estimate for $\eta$, although predictions are hindered by the 
arbitrariness of the employed coupling parameters. Actually, 
at the present stage no definite conclusion can be drawn about the shear viscosity of 
QGP: several microscopic models are yielding estimates of $\eta/s$, but no calculation 
still exists which takes into account all the relevant degrees of freedom.
Moreover further indications are needed both from experiment, beyond the information provided by $v_2$~\cite{Gavin}, as well as from 
hydro-calculations~\cite{Cassing,Romatschke}. Also the behaviour of this quantity in 
the hadronic phase~\cite{Chen1,Chen2} deserves further investigation in order to get 
quantitative information on the deconfinement phase transition.

{\bf Acknowledgements} The authors are grateful to D.~Davesne, for kindly providing the
calculation of $\eta/s$ in the pion gas; we also ackowledge fruitful discussions with 
P.~Czerski and A.~De Pace.

\end{document}